\documentclass[
  aps,prc,twocolumn,
  10pt,
  floatfix          
]{revtex4-2}
\usepackage{nameref}

\RequirePackage{lineno} 

\usepackage{silence}
\usepackage{graphicx}   
\usepackage{subfigure}  
\usepackage{placeins}   

\usepackage{xcolor}     
\usepackage[unicode, pdfencoding=auto]{hyperref}
\usepackage{natbib}
\usepackage[T1]{fontenc}          
\usepackage{textcomp}             

\usepackage{amsmath, amssymb, amsfonts}
\usepackage{xspace} 

\newcommand{\sqrtsNN} {\ensuremath{\sqrt{s_{\mathrm{NN}}}}\xspace}
\newcommand{\sevenn}        {$\sqrt{s_{\mathrm{NN}}}~=~7$~TeV\xspace}

\newcommand{\mpt}{\ensuremath{\langle p_{\rm T} \rangle}\xspace}
\newcommand{\ppt}{\ensuremath{p_{\rm T}}\xspace}
\newcommand{\mmt}{\ensuremath{m_{\rm T}}\xspace}

\newcommand {\auau}  {\ensuremath{\rm Au+Au}\xspace}
\newcommand {\pbpb}  {\ensuremath{\rm Pb+Pb}\xspace}

\newcommand {\oo}  {\ensuremath{\rm O+O}\xspace}
\newcommand {\pp}    {\ensuremath{\rm pp}\xspace}
\newcommand {\ppb}    {\ensuremath{\rm p+Pb}\xspace}

\newcommand {\dndeta}       {\ensuremath{\mathrm{d}N_\mathrm{ch}/\mathrm{d}\eta}\xspace}

\newcommand{\pythia}{\textsc{Pythia}}

\usepackage[bottom]{footmisc}

\begin{document}


\title{Predictions for Identified Hadron ($\pi^\pm$, $K^\pm$ and $p(\overline{p})$) Production and Collective Dynamics in Oxygen–Oxygen Collisions at $\sqrt{s_{NN}}$= 7 TeV with EPOS4, AMPT-SM, and Angantyr in \pythia~8}

\author{Rabia~Bashir}
\affiliation{Department of Physics, University of the Punjab, Lahore 54590, Pakistan}%

\author{Ramoona~Shehzadi}
\affiliation{Department of Physics, University of the Punjab, Lahore 54590, Pakistan}%

\author{M.~U.~Ashraf} 

\email{muashraf@wayne.edu; (Contact Author)}
\affiliation{Department of Physics and Astronomy, Wayne State University, 666 W. Hancock, Detroit, Michigan 48201, USA}%

\author{A.~M.~Khan}
\affiliation{Georgia State University, Atlanta, GA 30303, USA}%

\date{\today}

\begin{abstract}
We study the dynamics of identified hadrons ($\pi^\pm$, $K^\pm$ and $p(\overline{p})$) production in \oo collisions at \sevenn using recently updated version of EPOS4, string melting version of A Multi-Phase Transport Model (AMPT-SM) and Angantyr model, incorporated within \pythia~8. We examine the interplay between different mechanisms implemented in these models. Predictions for charged particle multiplicity ($dN_{ch}/d\eta$), transverse momentum ($p_T$) spectra of identified hadrons, particle yield ($dN/dy$) and mean transverse mass ($\langle m_T \rangle$) are presented. To probe the collective behavior of the produced particles, the $p_T$-differential kaons-to-pion and proton-to-pion ratios are studied. While AMPT incorporates some flow effects, EPOS4's implementation of full hydrodynamic flow proves significantly more effective. In contrast, the flow effects in $\textsc{Pythia}$~8 are substantially weaker compared to the other models. The upcoming $\ensuremath{\rm O+O}\xspace$ data from the LHC will help constrain the parameters of these models.

\end{abstract}

\maketitle

\section{Introduction}
\label{sec1}

Ultra-relativistic heavy-ion collisions at the Relativistic Heavy Ion Collider (RHIC) and the Large Hadron Collider (LHC) can create a new primordial state of matter known as the Quark-Gluon Plasma (QGP) characterized by the deconfinement of quarks and gluons. A wide range of experiments have been conducted using various colliding systems and different center-of-mass energies to understand the properties of the QGP, a dense and hot form of matter described by Quantum Chromodynamics (QCD). In this phase, quarks and gluons are the relevant degrees of freedom, unlike mesons and baryons, confined to color-neutral states~\cite{Busza:2018rrf, Becattini:2014rea, Aoki:2006we}. The hydrodynamic nature of the QGP, where it behaves like a perfect fluid~\cite{Heinz:2013th}, has been observed in symmetric heavy-ion ($A+A$) collisions, such as \pbpb and \auau at the LHC and RHIC respectively. On the other hand, proton-proton (\pp) collisions serve as a baseline for comparison.

Recent data from the LHC~\cite{ALICE:2016fzo, CMS:2016fnw} reveal compelling evidences for the formation of a QGP-like behavior in smaller collision systems. Observations in both, high multipilicty \pp and \ppb collisions\cite{Li:2012hc, ALICE:2013snk} exhibit collective flow phenomena traditionally associated with larger heavy-ion collisions. The observation of ``ridge'' structure long-range azimuthal correlations in high-multiplicity \pp collisions at the LHC~\cite{Werner:2010ss, Werner:2010ny} suggests the presence of hydrodynamic transverse flow, sparking speculation about the creation of small-scale the QGP droplets. This finding challenges the earlier assumption that small systems lack the extreme energy densities required for QGP formation. Investigating the mechanisms behind the QGP formation in these systems represents a pivotal question in high-energy QCD. These findings has generated substantial debate within the heavy-ion physics community and has significantly influenced interpretations of collision data. Consequently, exploring small system collisions at LHC energies is important for advancing our understanding of QGP formation and its underlying dynamics.

The LHC is anticipated to collide oxygen-oxygen (\oo) at $\sqrt{s_{NN}}$ $\approx$ 7\;TeV~\cite{Brewer:2021kiv}. This intermediate collision system aims to bridge the gap in particle multiplicity between smaller (\pp and \ppb) and \ppb) and larger (\pbpb) collisions, allowing for the study of how QGP signatures scale gradually with system size. Although \oo collisions involve a comparable number of nucleons to \ppb collisions, they exhibit a significantly less dense distribution of participants in the transverse plane, creating a distinct initial state that may influence the subsequent evolution of the system. A comprehensive analysis of oxygen nuclei can enhance our understanding of QGP-like features in small systems.  A series of recent theoretical studies~\cite{Rybczynski:2019adt, Lim:2018huo, Sievert:2019zjr, Huang:2019tgz, Schenke:2020mbo, Zakharov:2021uza, Huss:2020whe, Khan:2024fef, Ashraf:2024ocb} have investigated various aspects of particle production mechanisms, collective flow effects, and light nuclei production across different multiplicity ranges in \oo collisions.

Charged particle multiplicities (\dndeta), \ppt spectra, and ratios, collectively referred to as bulk observables and are fundamental probes for investigating the potential formation of the QGP in $A+A$ collisions. Correlations between particle multiplicity and transverse momentum provide insights into the properties of hot hadronic matter~\cite{VanHove:1982vk}, enabling the differentiation of soft and hard scattering processes.

This study presents a comparative analysis of identified hadron ($\pi^{\pm}, K^{\pm}, p(\bar{p})$) production at mid-rapidity ($|y| < 0.5$) in \oo collisions at $\sqrt{s_{NN}}$ = 7\;TeV, using the AMPT~\cite{Lin:2004en, Ma:2016fve}, \pythia~\cite{Bierlich:2018xfw}, and EPOS4~\cite{Porteboeuf:2010um, Werner:2023fne, Werner:2023zvo, Werner:2023jps, Werner:2023mod, Werner:2024ntd} event generators. The analysis focuses on global observables, including Bjorken energy density ($\epsilon_{Bj}$) , \ppt spectra, rapidity densities (dN/dy), and particle ratios.

The paper is structured as follows: Section~\ref{sec1} presents background on the research motivation and highlights the importance of \oo collisions. Section~\ref{sec2} outlines the event generation methodology using AMPT, \pythia and EPOS4 models. The results are presented and analyzed in Section~\ref{sec3}. Finally, Section~\ref{sec4} provides a summary of the study's key findings.

\section{Event Generators}
\label{sec2}

In this section, we present a brief overview of the AMPT, \pythia and EPOS4 models.

\subsection{AMPT}
A Multi Phase Transport Model (AMPT) is a transport model which has been extensively used to study the heavy-ion collisions at RHIC and LHC energies~\cite{Lin:2004en, Ma:2016fve}. The production of hadrons in AMPT is handled by the HIJING~\cite{Wang:1991hta}, incorporating two key components: the initial spatial and momentum distributions of minijet partons, and the soft string excitations. The ZPC parton cascade model~\cite{Zhang:1997ej} subsequently governs the progression of these partons through space and time. After the evolution of parton cascade, the model converts the remaining partonic degrees of freedom into final-state hadrons using either quark coalescence or string fragmentation. Following their formation, the interactions of these hadrons are modeled by the Hadronic Transport Model (ART)~\cite{Li:1995pra}. Two versions of AMPT model, namely, the default AMPT, and string melting (AMPT-SM) are available for simulations. The default AMPT, which relies only on minijet partons from HIJING in the parton cascade and Lund string fragmentation for hadronization~\cite{Sjostrand:2007gs}, can  reasonably reproduce the \ppt spectra and rapidity distributions of identified particles at SPS and RHIC energies, however it underestimates the elliptic flow at RHIC energies~\cite{Lin:2001zk}. In the AMPT-SM model~\cite{Lin:2001zk}, initially, Lund string fragmentation  produces hadrons, which are then decomposed into their valence quarks. These quarks are later recombined into hadrons via a simple quark coalescence model after the ZPC. Successful descriptions of anisotropic flows have been achieved in both large and small collision systems~\cite{Lin:2001zk, Bzdak:2014dia, Ma:2014pva} with this approach.

In this analysis, we used the string melting version of the AMPT model to simulate a dataset of $\sim 2.5$ million minimum-bias $\ensuremath{\rm O+O}\xspace$ events at ${\sqrt{s_{NN}}}$ = 7\;TeV.

\subsection{\pythia}

Angantyr~\cite{Bierlich:2018xfw}, a \pythia~8 based event generator, has been developed to collide heavy nuclei like $p-A$ and $A-A$ collisions. It incorporates an advanced Glauber model inspired by the color fluctuation model~\cite{Heiselberg:1991is, Blaettel:1993ah, Alvioli:2013vk, Alvioli:2014sba, Alvioli:2014eda}, along with the DIPSY generator~\cite{Avsar:2005iz, Avsar:2006jy, Flensburg:2011kk} and \pythia~8's \pp machinery~\cite{Sjostrand:2014zea}. The color fluctuation model has been enhanced by accounting fluctuations not only in the $N-N$ cross-section but also in individual nucleons. This enables the application of the same model to both $p-A$ and $A-A$ collisions, allowing for the classification of each individual $N-N$ sub-collisions as diffractive or (primary or secondary) non-diffractive collisions. \pythia~8 generates corresponding sub-collisions with a specialized diffractive-like treatment for secondary non-diffractive collisions. These sub-collisions are then stacked into full $p-A$ or $A-A$ events using a model inspired by Fritiof~\cite{Andersson:1986gw} and the wounded nucleon model~\cite{Bialas:1976ed}, using \pythia's MPI machinery. This approach extrapolates \pp collision dynamics to heavy-ion collisions using a simplified framework, assuming no collective effects between the constituent sub-collisions. Despite its simplicity, the model accurately predicts charged particle rapidity distributions in $p-A$ collisions using minimal adjustable parameters~\cite{Bierlich:2018xfw}, and subsequently provides a reliable description of multiplicity distributions in $A-A$ collisions~\cite{Bierlich:2018xfw, ALICE:2018cpu} without any additional calibration.

For the current study, $\approx$ 5 million events were generated using Angantyr tune of \pythia.

\subsection{EPOS4}
EPOS is a comprehensive simulation framework that employs a quantum mechanical multiple scattering approach to describe high-energy collisions. By integrating parton ladders, off-shell remnants, and saturation of parton ladders within a Monte Carlo framework, EPOS provides a detailed understanding of both the initial and final stages of \pp and $A+A$ collisions.
EPOS uses a combined Gribov-Regge theory and eikonalized parton model for the treatment of the first interactions after collisions. This approach ensures conservation laws and treats subsequent interactions equally~\cite{Drescher:2000ha}. Particle production is calculated using Feynman diagrams of QCD-inspired effective field theory. Nucleons consist of constituents with momentum fractions that add up to one. Nucleons can be ``spectator'', ``participants'', or ``remnants''. Spectators don't interact, while participants or remnants do. EPOS uses its own string model for particle production, which is similar to the Lund string model but has some key differences ~\cite{Werner:1993uh}.

\begin{table*}[!]
    \centering
    \caption{ Charged-particle multiplicity ($\langle\mathrm{d}N_\mathrm{ch}/\mathrm{d}\eta\rangle$) values at $|\eta| < 0.5$ in \oo collisions at $\sqrt{s_{NN}}$= 7\;TeV. The values are shown for different centrality classes using \pythia~8, AMPT-SM and EPOS4.}
    \begin{tabular}{cccc}
    \hline
    \hline
    Centrality ($\%$) &  \pythia~8 & AMPT-SM &  EPOS4\\ 
     & $\ensuremath{\langle\mathrm{d}N_\mathrm{ch}/\mathrm{d}\eta}\rangle\xspace\pm{\rm rms}$  & $\ensuremath{\langle\mathrm{d}N_\mathrm{ch}/\mathrm{d}\eta}\rangle\xspace\pm{\rm rms}$ & $\ensuremath{\langle\mathrm{d}N_\mathrm{ch}/\mathrm{d}\eta}\rangle\xspace\pm{\rm rms}$ \\
    \hline   
            $0$~--~$5$    & $151.934\pm0.028$ & $188.293\pm0.044$ & $236.898\pm0.064$ \\
            $5$~--~$10$   & $122.193\pm0.025$ & $145.678\pm0.038$ & $189.679\pm0.057$ \\
            $10$~--~$20$  & $97.452\pm0.016$  & $110.442\pm0.024$ & $105.717\pm0.030$ \\
            $20$~--~$30$  & $72.190\pm0.013$  & $76.085\pm0.020$  & $72.190\pm0.013$  \\ 
            $30$~--~$40$  & $52.405\pm0.011$  & $51.471\pm0.016$  & $75.086\pm0.026$  \\ 
            $40$~--~$50$  & $36.969\pm0.010$  & $34.188\pm0.013$  & $53.145\pm0.021$  \\ 
            $50$~--~$60$  & $25.127\pm0.008$  & $22.082\pm0.011$  & $37.145\pm0.018$  \\ 
            $60$~--~$80$  & $12.565\pm0.004$  & $11.117\pm0.005$  & $19.875\pm.009$   \\ 
            $80$~--~$100$ & $3.254\pm0.002$   & $3.761\pm0.003$   & $5.118\pm0.005$   \\ 
    \hline
    \hline
    \end{tabular}
    \label{tab1}
\end{table*}

EPOS introduces a dynamic process of division of string segments into the ``core'' and the ``corona'' regions~\cite{Werner:2007bf, Werner:2010aa, Werner:2013tya} to address the issue of high string density in high multiplicity \pp and $A+A$ collisions, where individual strings cannot decay independently. The model combines Regge theory~\cite{Werner:2019aqa} for low-density regions (corona) and hydrodynamic equations for high-density regions (core) to simulate particle production across varying densities. A new version of EPOS, known as EPOS4 has been recently released~\cite{Werner:2023fne, Werner:2023zvo, Werner:2023jps, Werner:2023mod}. EPOS4 introduces a novel concept that offers a deeper understanding of the fundamental relationships between four main principles in \pp and heavy-ion collisions: rigorous parallel scattering, energy conservation, factorization, and saturation~\cite{Werner:2023mod}.
This framework accurately models high-\ppt{} particle production and collective effects in high-multiplicity events, while the dynamic saturation scale implementation does not affect high-\ppt{} particle production even with multiple parallel scatterings~\cite{Werner:2023zvo}. For more information, see Refs.~\cite{Werner:2023fne, Werner:2023zvo, Werner:2023jps, Werner:2023mod}. 

Using EPOS4, we generated $\sim 1.5$ million minimum bias \oo events, at ${\sqrt{s_{NN}}}$ = 7\;TeV. We ran EPOS4 simulations with specific parameters for the \oo system. We enabled core-corona effects, hydrodynamic evolution, and hadronic cascade to model collective flow and particle interactions. The centrality classes were determined using the charged particle multiplicy at $|y| < 0.5$ and is listed in Table.~\ref{tab1}.

\begin{figure*}[ht!]
  \centering
  \includegraphics[width=0.9\textwidth]{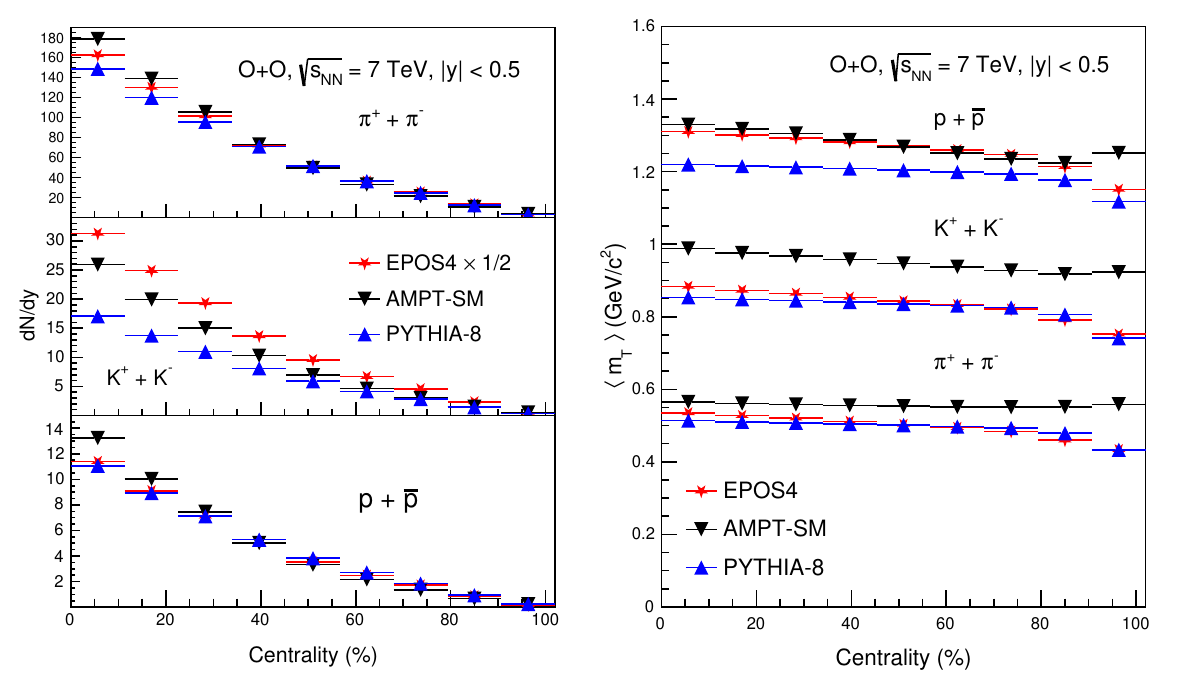}
  \caption{(Color online) The centrality dependence of integrated yields ($dN/dy$) of (left) and $\langle m_T \rangle$ of pions, kaons and protons at $|y|< 0.5$ in \oo collisions at $\sqrt{s_{NN}}$= 7\;TeV from  AMPT-SM EPOS4, and \pythia~8. EPOS4 results are scaled by $1/2$.}
  \label{fig1}
\end{figure*}

\section{Results and Discussion}
\label{sec3}

In this section, the predictions for identified charged hadron ($\pi^\pm, K^\pm, p({\overline{p}})$) production at $|y|< 0.5$ in \oo collisions at $\sqrt{s_{NN}}$ = 7\;TeV for different centrality classes using \pythia~8, AMPT-SM, and EPOS4 are presented. Hereafter, $\pi^{+} + \pi^{-}, K^{+} + K^{-}, p + \bar{p}$ will be referred to as pions, kaons, and protons, respectively and will be used in the text throuout.



\subsection{Bjorken energy density}
The experimental determination of the initial energy density produced in $A+A$ collisions is a critical objective for characterizing the QGP properties. This quantity serves as a fundamental thermodynamic parameter, constraining the equation of state and transport coefficients of the hot, dense matter formed. The initial energy density is commonly estimated using the Bjorken boost-invariant hydrodynamic model~\cite{Bjorken:1982qr}. This model relates the measured transverse energy density ($E_T$) at mid-rapidity to the initial energy density, assuming longitudinal boost invariance. The quantitative estimation is obtained by applying the Bjorken formula, typically expressed as
\begin{equation}
    \epsilon_{Bj} = \frac{1}{\tau S_{T}} \frac{dE_T}{dy}
\end{equation}\label{eq1}

where the transverse overlap area of the colliding nuclei is represented by $S_{T}$, and $dE_T/dy$ denotes the transverse energy density at mid-rapidity at a formation time $\tau$. Equation~\ref{eq1} exhibits a divergence as the formation time approaches zero. Consequently, for the calculation of the Bjorken energy density in this analysis, a finite formation time of $\tau = 1$ is assumed. This regularization is necessary to avoid non-physical infinities in the estimation of the initial energy density. The total transverse energy produced in an event is denoted as $E_T$, and the transverse overlap area of the colliding nuclei is defined as $S_T = \pi R^2$, where $R$ represents the effective nuclear radius. Replacing, $R = R_0 A^{1/3}$ and $A= N_{part}/2$,

\begin{equation}
    S_T = \pi R_0^2 \left (\frac{N_{part}}{2}\right) ^{2/3}
\end{equation}\label{eq2}

Given the dominant contribution of pions, kaons, and protons to the total transverse energy due to their high production yields, the total transverse energy ($E_T$) can be accurately approximated by considering these particle species~\cite{ALICE:2016igk, STAR:2008med, Karsch:2001vs}.

\begin{equation}
    \frac{dE_T}{dy} \approx \frac{3}{2} \times \left ( \langle m_T \rangle \frac{dN}{dy} \right)_{\pi^\pm} + 2 \times \left( \langle m_T \rangle \frac{dN}{dy} \right)_{K^\pm, p, \overline{p}}
\end{equation}\label{eq3}

Here, $m_T = \sqrt{p_{T}^2 + m^2}$ is transverse mass and $dN/dy$ is the integrated yield of pions, kaons and protons at the $|y|< 0.5$. The multiplicative factor in each term serves as a correction for the contributions of corresponding neutral particles, which are not directly measured in this analysis. Simplified form of Eq.~\ref{eq3} is;

\begin{equation}\label{eq4}
\begin{split}
\frac{dE_T}{dy} \approx {} & 
\frac{1}{\tau \,\pi R_0^2 \bigl(\tfrac{N_{part}}{2}\bigr)^{2/3}}
\Biggl[
    \frac{3}{2}\,\Bigl(\langle m_T \rangle \tfrac{dN}{dy}\Bigr)_{\pi^\pm} \\
&\quad
    +\;2\,\Bigl(\langle m_T \rangle \tfrac{dN}{dy}\Bigr)_{K^\pm,\,p(\overline{p})}
\Biggr]
\end{split}
\end{equation}

The centrality dependence of integrated particle yields ($dN/dy$) at mid-rapidity for pions, kaons, and protons in \oo collisions at $\sqrt{s_{NN}}$= 7\;TeV is illustrated in Fig.~\ref{fig1} (left). A comparative study of EPOS4, AMPT-SM, and \pythia~8 predictions reveals a universal trend of decreasing yields with increasing centrality. Notably, EPOS4 consistently predicts the highest yields, and \pythia~8 the lowest. The observed hierarchy of yields, with pions being the most abundant, followed by kaons and protons, aligns with thermalized Boltzmann production mechanisms in high-energy nuclear reactions. The increased kaon production observed in EPOS4 simulations is attributed to the hydrodynamic formation of a dense medium, which subsequently facilitates strange quark production via in-medium processes.

Figure~\ref{fig1} (right) presents the mean transverse mass ($\langle m_T \rangle$) as a function of centrality in \oo collisions at $\sqrt{s_{NN}}$= 7\;TeV from EPOS4, AMPT-SM and \pythia~8. It is observed that \mmt exhibits a weak dependence on centrality across all models. Protons have the highest \mmt, followed by kaons, and pions, which is expected due to their mass ordering, $m_p>m_K>m_\pi$ which is consistent with the expectation that heavier particles have higher average transverse mass due to their mass-dependent momentum distributions. AMPT-SM predicts higher \mmt for all particles, especially for protons and kaons. AMPT-SM includes strong partonic interactions and hadron re-scattering, leading to higher \mmt values due to enhanced transverse flow. The decreasing trend of \mmt with centrality suggests stronger final-state interactions in central collisions. EPOS4 predicts \mmt, lying between AMPT-SM and \pythia~8. EPOS4 is hydrodynamic model which includes transverse flow from both partonic and hadronic phases, and the flow is influenced by multi-parton interactions and flow-induced correlations. On the other hand, \pythia~8, which lacks both partonic interactions and a hydrodynamic phase, shows the lowest \mmt, consistent with a purely perturbative QCD-based approach. The \mmt remains nearly the same, indicating a similar spectral shape for all models.

\begin{figure}[!]
    \centering
    \includegraphics[width=\columnwidth]{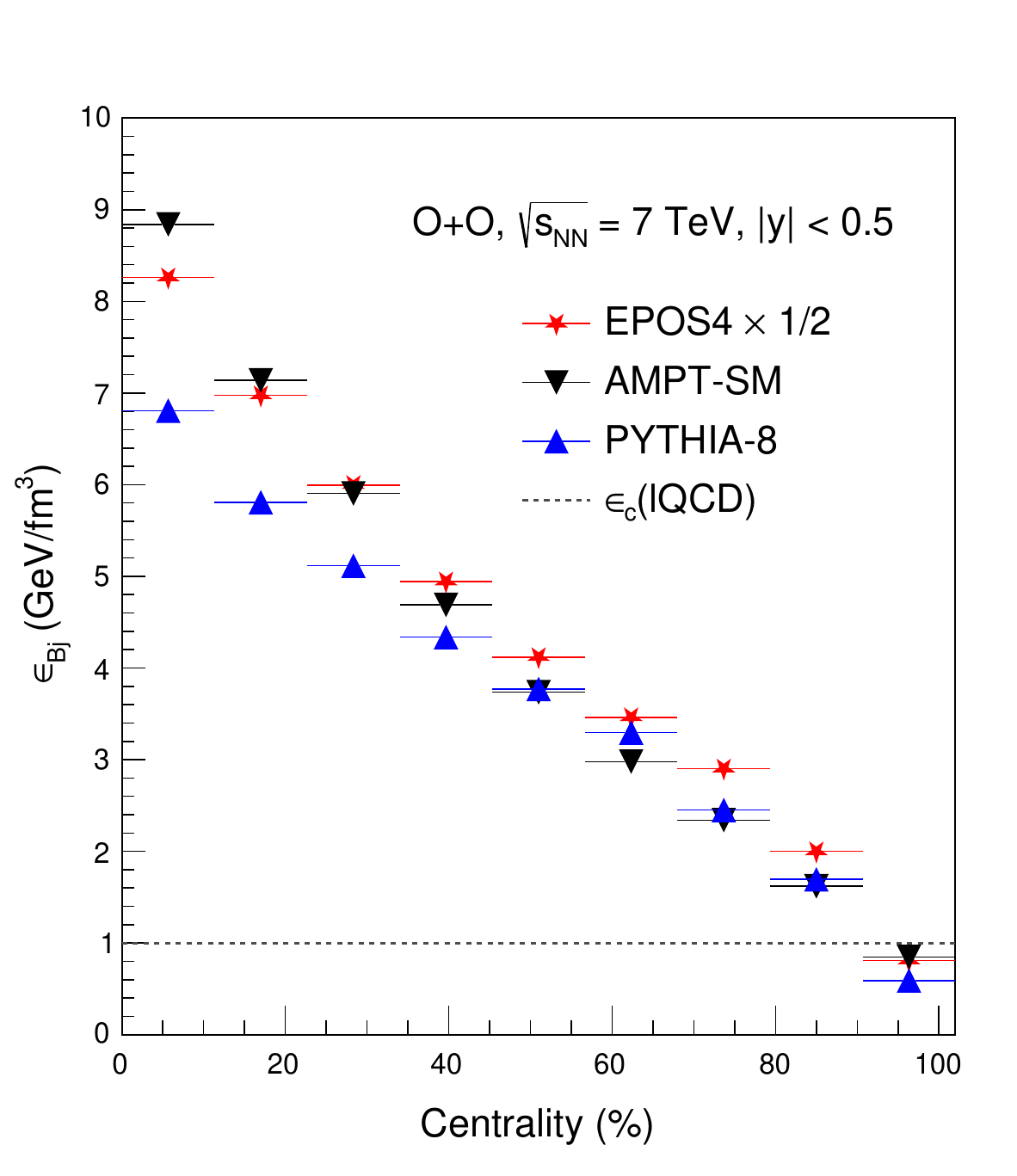}%
    \caption{(Color online) Centrality dependence of Bjorken energy density ($\varepsilon_{Bj}$) at $|y|< 0.5$ in \oo collisions at $\sqrt{s_{NN}}= 7\;$TeV from EPOS4, AMPT-SM, and \pythia. Different symbols show various models. EPOS4 simulations are scaled by a factor of 0.5.}
    \label{fig2}
\end{figure}


\begin{figure*}[t!] 
    \centering 

    \includegraphics[width=0.33\textwidth]{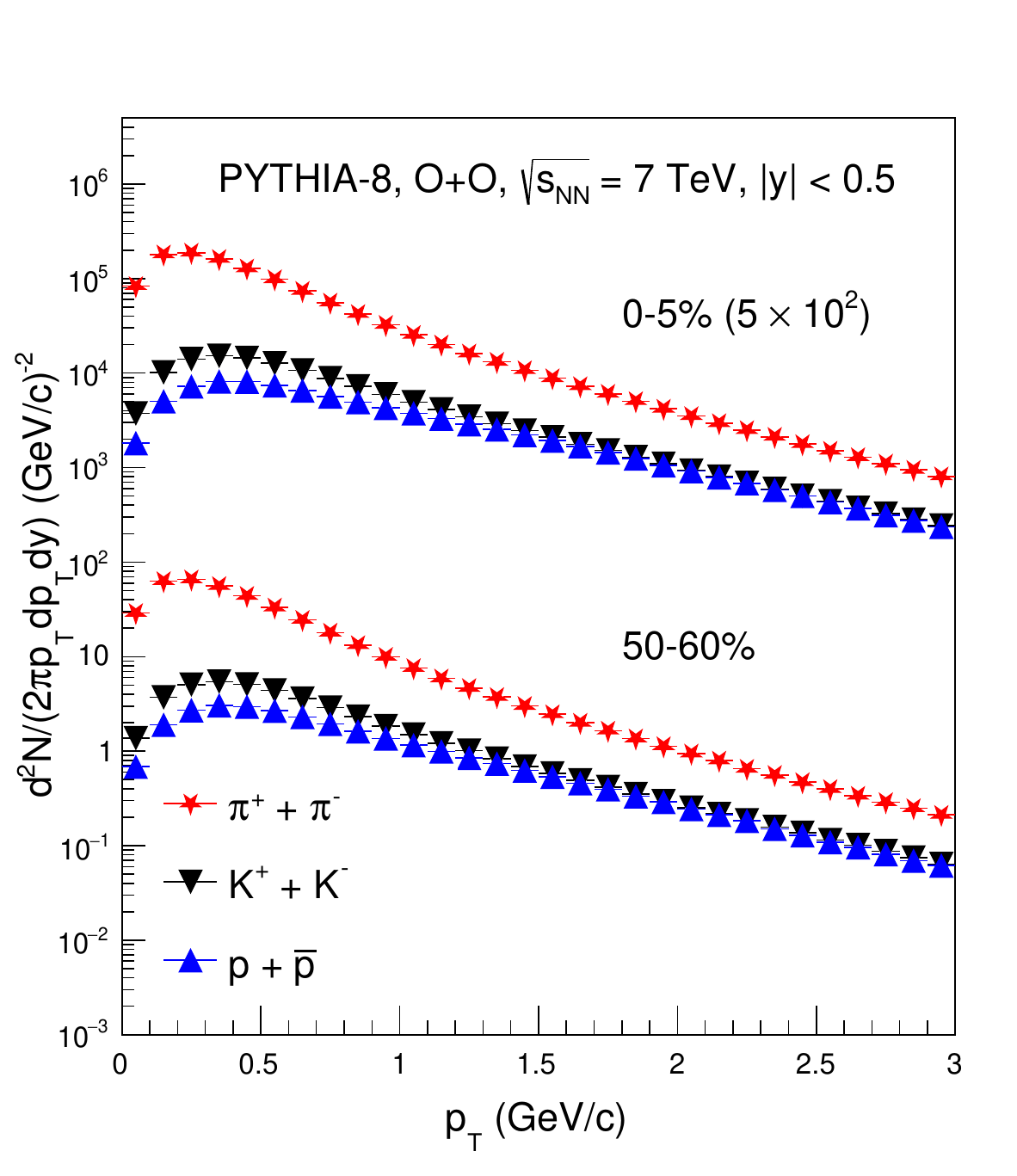}%
    \includegraphics[width=0.33\textwidth]{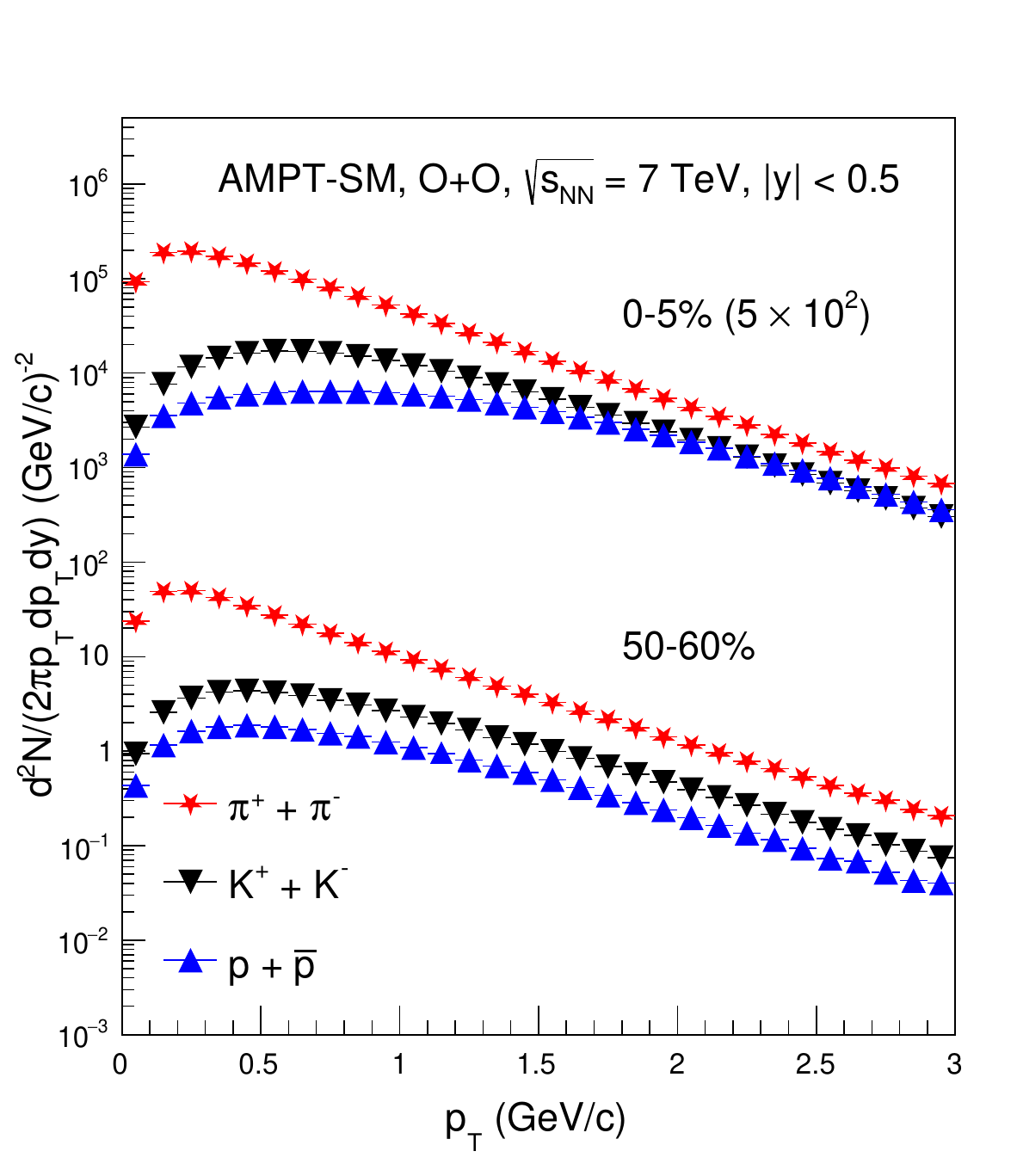}%
    \includegraphics[width=0.33\textwidth]{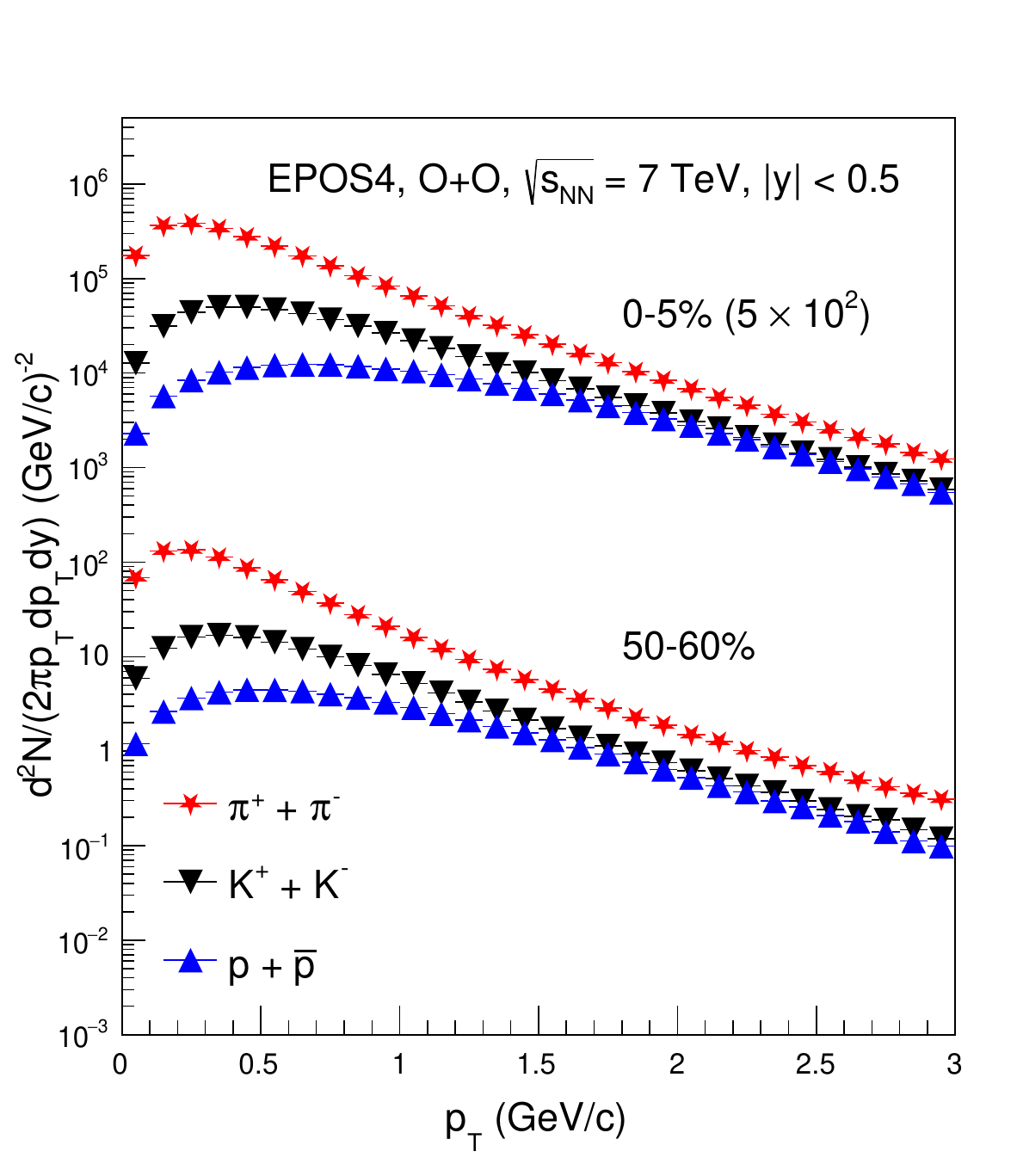}%

    \caption{(Color online) The transverse momentum {\ppt} spectra of pions, kaons and protons at $|y|< 0.5$ in central (0–5$\%$) and peripheral (50–60$\%$) \oo collisions at $\sqrt{s_{NN}}$= 7\;TeV from \pythia{}/Angantyar (left), AMPT-SM (middle) and EPOS4 (right). Markers of different style shows the prediction from different models. }
    \label{fig3}
\end{figure*}


Figure~\ref{fig2} presents the Bjorken energy density ($\varepsilon_{Bj}$) extracted using Eq.~\ref{eq4} at $|y|< 0.5$ as a function of centrality in \oo collisions at $\sqrt{s_{NN}}$= 7\;TeV comparing predictions from EPOS4, AMPT-SM, and \pythia~8. The general trend shows that $\varepsilon_{Bj}$ is highest in the most central collisions and decreases toward peripheral collisions. This behavior aligns with expectations, as more central collisions involve a greater overlap of nuclei, leading to higher particle density and energy deposition per unit volume. It is evident from Eq.~\ref{eq4} that $\varepsilon_{Bj}$ is strongly dependent on the multiplicity. It is observed that EPOS4 predicts the highest energy density across all centralities which attributed to its inclusion of hydrodynamic evolution and multi-parton interactions. These features enhance collectivity and thermalization, leading to a scenario consistent with the QGP formation and high multiplicity in most central collisions. AMPT-SM predicts lower energy densities than EPOS4 which is due to the difference in the initial states conditions in both of the models. The string-melting mechanism in AMPT allows for some level of partonic interactions before hadronization, resulting in moderate collective effects, though not as pronounced as in EPOS4. The lower $\varepsilon_{Bj}$ compared to EPOS4 suggests that AMPT does not generate as strong an initial energy density, possibly due to differences in parton scattering rates or equation of state assumptions. On the other hand, \pythia~8 predicts the lowest energy densities, as it is a purely hadronic model that does not incorporate QGP-like effects or significant collective behavior. The dashed line in fig.~\ref{fig2} representing a critical energy density threshold $\varepsilon_{c}$, indicating the QGP formation threshold from lattice QCD calculations~\cite{Karsch:2001vs}. It is observed that all the models are above the threshold value which hints at observing the signals of QGP in \oo collisions at the LHC energies. 

\subsection{Transverse momentum ($p_T$) spectra}
The transverse momentum {\ppt} spectra of identified particles are essential for understanding particle production, QGP dynamics, and hadronization in heavy-ion collisions~\cite{Andronic:2014zha, Andronic:2017pug, Andronic:2020iyg}. Fig.~\ref{fig3} compares the transverse momentum (\ppt) spectra of identified hadrons at $|y|< 0.5$ in \oo collisions at $\sqrt{s}$= 7 TeV from \pythia~8 (left), AMPT-SM (middle) and EPOS4 (right). The spectra are shown for two centrality classes; (0–5$\%$) central collisions are scaled by a factor of $5\times10^{2}$ for better visualization and (50-60\%) peripheral collisions. Pions exhibit the highest yields across all models and centralities, followed by kaons and protons, reflecting the particle mass hierarchy. Central collisions show significantly higher yields compared to peripheral collisions, and generally exhibit a flatter slope at higher \ppt, indicating a higher average transverse momentum. The slope of the \ppt spectra predicted by EPOS4 compared  to other models is relatively flat, particularly for protons, suggesting stronger radial flow~\cite{Ashraf:2024ocb}. This is consistent with EPOS4 incorporating hydrodynamic evolution, which enhances the collective expansion of the medium and pushes heavier particles to higher \ppt. AMPT-SM also shows evidence of collective flow due to partonic scatterings in the string-melting mechanism, but the effect is less pronounced compared to EPOS4. \pythia~8, on the other hand, predicts the softest spectra, particularly for protons, as it lacks partonic interactions and collective expansion, relying primarily on hadronization from string fragmentation. The difference between the central (0-5\%) and mid-peripheral (50-60\%) collisions is also significant. In all simulations, the spectra in central collisions are systematically higher due to the larger number of participating nucleons. Additionally, the convergence of the \ppt spectra at intermediate \ppt across all models suggests the presence of radial flow, which is known to push heavier particles to higher \ppt. This effect is strongest in EPOS4, moderate in AMPT-SM, and weakest in PYTHIA-8.


\begin{figure*} 
    \centering 

    \includegraphics[width=8cm]{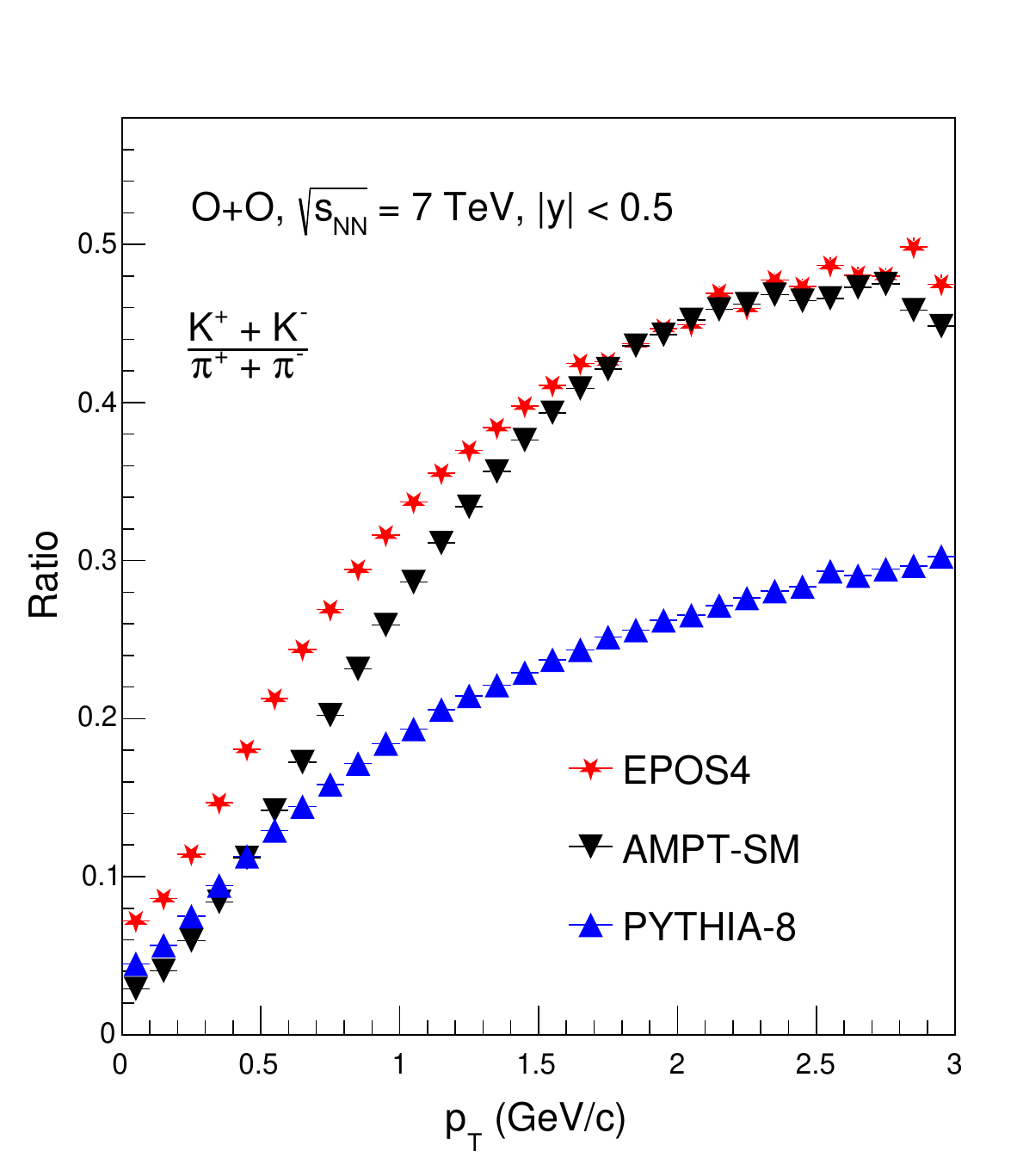}%
    \includegraphics[width=8cm]{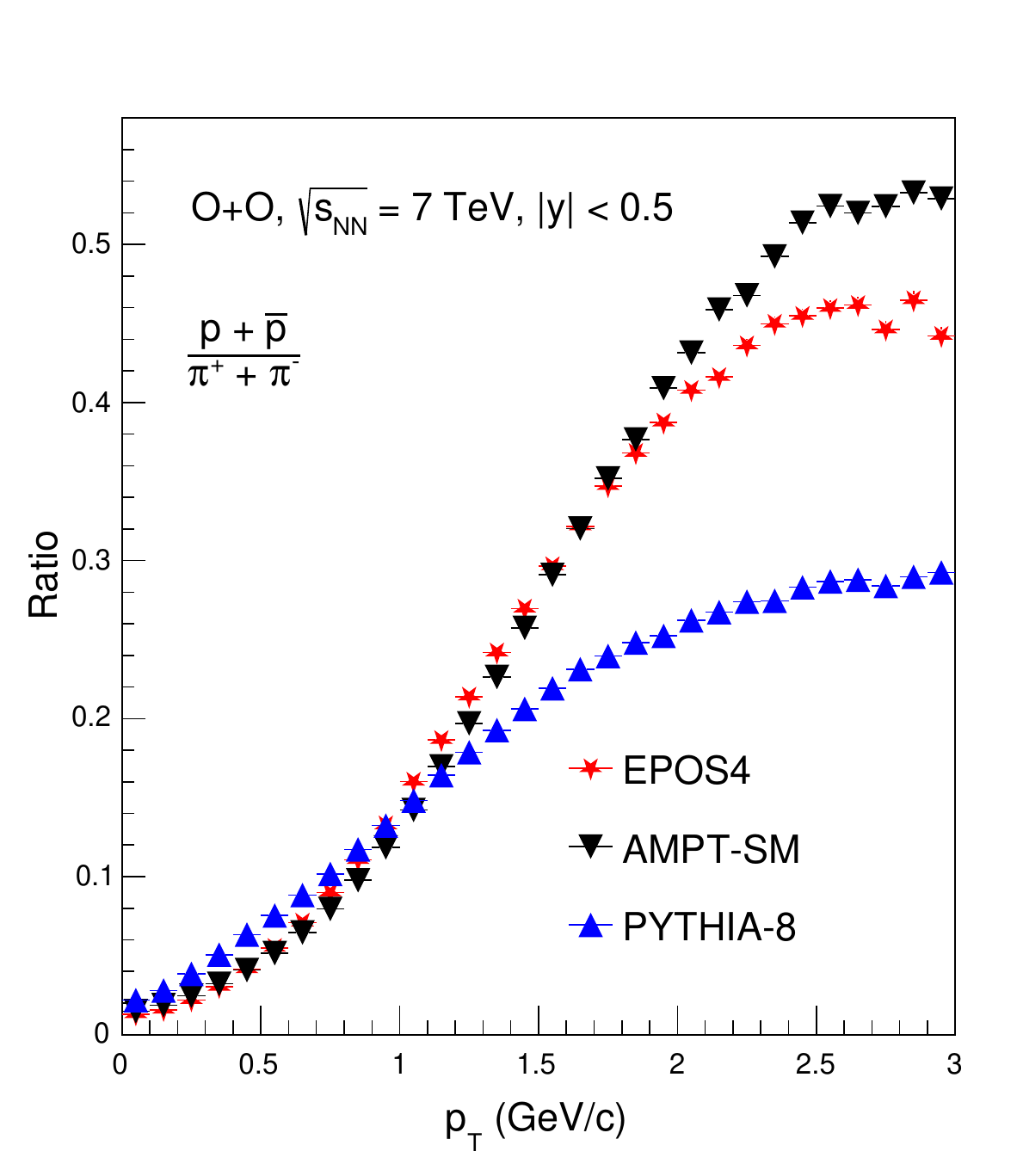}%

    \caption{(Color online) \ppt-differential kaon-to-pion ratio (left) and proton-to-pion ratio (right ) at mid-rapidity in central (0–5$\%$) \oo collisions at $\sqrt{s_{NN}}$= 7\;TeV from \pythia{}/Angantyar, AMPT-SM and EPOS4. Markers of different style shows the prediction from different models. }
    \label{fig4}
\end{figure*}


\subsection{Particle ratios}

This section presents predictions of \ppt-differential ratios for kaons and protons relative to pions in \oo collisions at $\sqrt{s_{NN}}$= 7\;TeV, using EPOS4, AMPT, and \pythia~8. Particle ratios provide insights into the relative production of different hadron species, which can reveal information about the underlying production mechanisms and the properties of the medium created in the collisions. Figure~\ref{fig4} presents a comparison of particle yield ratios as a function of transverse momentum \ppt in (0–5$\%$) central \oo collisions at $\sqrt{s_{NN}}$= 7\;TeV. The left panel shows the kaon-to-pion ratio, while the right panel displays the proton-to-pion ratio. Three different simulations EPOS4, AMPT-SM, and PYTHIA-8—are compared. The ratios increase with \ppt, indicating that kaons and protons become relatively more abundant at higher transverse momentum compared to pions. This trend is consistent with expectations from mass-dependent spectral shapes due to radial flow and hadronization dynamics. EPOS4 exhibits a higher kaon-to-pion and proton-to-pion ratio than PYTHIA-8, particularly at intermediate \ppt. This suggests that EPOS4 generates more pronounced radial flow or different hadronization dynamics. Additionally, EPOS4 is a hydrodynamic model that produces a QGP-like medium. This medium facilitates the production of strange quarks resulting in enhanced kaon-to-pion ratio at intermediate \ppt. The AMPT-SM model generally predicts ratios similar to EPOS4 but with slightly different trends at intermediate \ppt. AMPT includes partonic interactions and a hadron transport phase, which may explain the closer agreement with EPOS4. \pythia{}/Angantyar systematically predicts lower ratios compared to EPOS4 and AMPT-SM, especially at intermediate \ppt. Since \pythia~8 is a purely perturbative QCD-based model without partonic medium interactions, it tends to underestimate radial flow effects, leading to lower kaon and proton production relative to pions.

\subsection{Mean Transverse momentum ({\mpt})}

\begin{figure}[!]
    \centering
    \includegraphics[width=\columnwidth]{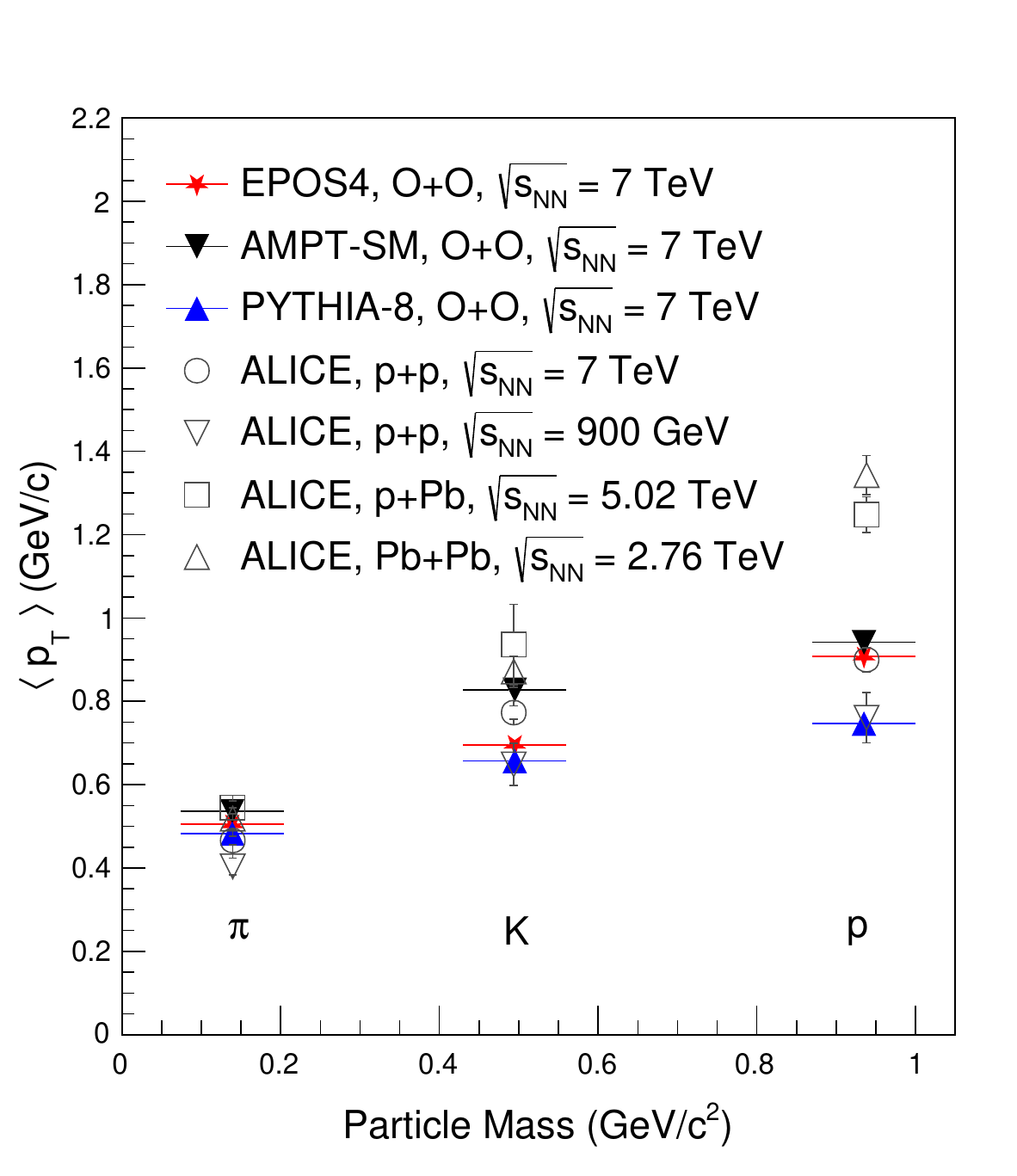}
    \caption{(Color online) Mean Transverse momentum {\mpt} of pions, kaons and protons at mid-rapidity as a function of particle masses in central (0–5$\%$) \oo collisions at $\sqrt{s_{NN}}$= 7\;TeV from EPOS4, AMPT-SM and \pythia. Different symbols show various models.}
    \label{fig5}
\end{figure}

The mean transverse momentum {\mpt} of pions, kaons and protons at $|y|< 0.5$ as a function of particle masses in central (0–5$\%$) \oo collisions at $\sqrt{s_{NN}}$= 7\;TeV from EPOS4, AMPT-S and \pythia~8 is shown in Fig.~\ref{fig5}. The predictions from the simulations were compared with the published data from \pp, \ppb and $Pb+Pb$ collisions from the ALICE experiment. It is observed that \mpt follows a mass hierarchy across different systems, reflecting the impact of collective expansion. Heavier particles receive a stronger boost from the collective expansion of the medium, leading to higher \mpt. Both, EPOS4 and AMPT-SM predict similar \mpt for each particle species, with AMPT-SM slightly higher, particularly for protons. This suggests that both models incorporate collective effects, but AMPT-SM may have stronger partonic interactions or final-state rescattering. These predictions from simulations in \oo  collisions are in reasonable agreement with experimental data from $pp$ collisions at \sqrtsNN = 900 GeV~\cite{42} and 7 TeV\cite{39a}, \ppb collisions at \sqrtsNN = 5.02 TeV~\cite{42a} and \pbpb collisions at \sqrtsNN = 2.76 TeV~\cite{41a} especially for pions and kaons. However, for protons, the ALICE \ppb data at higher energy shows a significantly higher \mpt, indicating that larger collision systems with more collective flow lead to stronger boosts for heavier particles. While none of the model explain the \mpt of protons. The upcoming \oo collisions at the LHC will be helpful to further constrain the model parameters for better understanding.

\section{Conclusions}\label{sec4}
This study investigates the production of various identified hadron ($\pi^\pm$, $K^\pm$ and $p(\overline{p})$) in \oo collisions at \sevenn using EPOS4, AMPT-SM and \pythia~8 Angantyr. The key findings of this investigation are summarized below:

\begin{itemize}
    \item The distribution of all charged-particle multiplicity (\dndeta) is successfully reproduced by all models used for the current study. However, EPOS4 predicts the higher multiplicity particularly in most central collisions where the core part is dominated. The comparison of these predictions with future experimental data from the LHC experiments will certainly be helpful to check the model’s capability of describing particle production in \oo collisions.
    \item All of the models successfully reproduce the shape of \ppt spectra for $\pi^\pm$, $K^\pm$ and $p(\overline{p})$. A clear mass ordering is observed, consistent with observations in other collision systems. The convergence of heavier particle (proton) spectra with lighter particle (pions) at intermediate \ppt is more pronounced in EPOS4 provides hint for the presence of radial flow in \oo collisions. 
    \item The difference in the $dN/dy$ of particle type is due to the interplay of different physics mechanisms between the models. $dN/dy$ exhibits a strongs centrality dependent for all models. 
    \item We observed a strong centrality dependence in $K/\pi$ and $p/\pi$ ratios especially in EPOS4 and AMPT-SM compared to \pythia~8. The radial flow effects in $p/\pi$ ratio in \pythia~8 are less significant. $K/\pi$ ratios exhibits an increase in strangeness enhancement with increasing \ppt in EPOS4 and AMPT-SM. This effect is less significant in \pythia~8.
    \item We observe an increase in average transverse momentum (\mpt) with increasing centrality indicating stronger radial flow in more central collisions. The models follow the trend established by the already existing data from different collision systems. 
\end{itemize}

To fully interpret these observations, a comparison with future experimental data is essential. The forthcoming data from \oo collisions at the LHC will be particularly valuable, offering critical insights into the heavy-ion-like behavior observed in small systems and enabling the refinement of model parameters.


\bibliographystyle{utphys}
\bibliography{bib}

\end{document}